\documentclass[epj]{svjour}
\usepackage[T1]{fontenc}
\usepackage{mathptmx}
\usepackage[scaled=0.92]{helvet}
\usepackage{amsmath,amssymb}
\usepackage{graphicx}
\usepackage{booktabs}
\usepackage[numbers,sort&compress]{natbib}
\usepackage{xcolor}
\usepackage{url}
\usepackage{microtype}
\usepackage[colorlinks=true,linkcolor=blue!50!black,citecolor=blue!50!black,urlcolor=blue!50!black]{hyperref}
\hypersetup{pdftitle={Data-quality monitoring of luminosity blocks with ATLAS Open Data: an open benchmark and a learned within-run detector},
  pdfauthor={Renjie Wang}}

\newcommand{\Ct}{\textsc{ctx}}
\newcommand{\Cs}{\textsc{single}}

\journalname{Eur. Phys. J. C}

\begin{document}

\title{Data-quality monitoring of luminosity blocks with ATLAS Open Data:\\
an open benchmark and a learned within-run detector}
\titlerunning{Data-quality monitoring of luminosity blocks with ATLAS Open Data}
\author{Renjie Wang\thanks{\emph{e-mail:} \href{mailto:rjwang@ihep.ac.cn}{\texttt{rjwang@ihep.ac.cn}}}}
\authorrunning{R. Wang}
\institute{Institute of High Energy Physics, Chinese Academy of Sciences, Beijing 100049, China}
\date{Date: \today}

\abstract{%
Data-quality monitoring at the LHC decides, luminosity block by luminosity block, which collision data
enter physics analyses, yet its methods cannot be compared openly because the experiments' monitoring
data are not public.
We construct a public benchmark from the ATLAS Open Data release of the 2015--2016 proton--proton
collisions at $\sqrt{s}=13$~TeV: 1289-bin histograms of reconstructed jets, electrons, photons, muons and
the missing transverse momentum computed from them, for 39\,828 luminosity blocks of all 217 runs in the
Good Run Lists; eight types of detector defects injected at event level at three strengths; real-data labels
from the Good Run Lists and the liquid-argon calorimeter error flag; and a time-ordered protocol in which
every detector is trained and calibrated on earlier runs only.
Histogram shapes differ far more between runs than within a run (median $\chi^2/n_\mathrm{dof}$ of 5.9
against 1.1), so references and autoencoders built across runs fail, and all competitive detectors use
residuals with respect to a robust reference of each run.
A gradient-boosted classifier on position-invariant features of these residuals, trained on synthetic
defects and calibrated by split-conformal prediction, detects 43\% of the luminosity blocks with injected
defects at a false-positive rate of 1.0\%, against 25\% for a matched-filter scan and 63\% for a
single-histogram test that knows the defect pattern.
For defect types excluded from its training, its efficiency falls to 18\%; adding pattern-agnostic
synthetic detector failures to the training raises it to 34\%, above the scan, at a cost of three
percentage points on the known types.
}

\maketitle

\section{Introduction}
\label{sec:intro}

ATLAS records its collision data in luminosity blocks (LBs), intervals of typically one minute within which
the instantaneous luminosity and the detector conditions are treated as constant; CMS uses luminosity
sections of about 23~s \cite{Pol:2019CMSDQM}. Whether an LB can be used in physics analyses is decided by
data-quality (DQ) monitoring and assessment: histograms of detector-level and reconstructed-object
quantities are filled for every run and compared with references by automatic checks and by shifters, and
problems are recorded as defects per LB, from which the Good Run Lists (GRLs) are derived. ATLAS describes
these procedures and their performance for the 2015--2018 data taking in Ref.~\cite{ATLAS:DQ2015-2018}.
Every LB that is wrongly rejected is lost integrated luminosity, and every defect that is missed
contaminates the analyses that use the data, so a DQ system operates at a false-alarm rate fixed in advance,
at which as many defects as possible should be found.

Machine-learning methods have been studied for this task in several experiments, as part of a wider use of
anomaly detection in particle physics \cite{Belis:2024review}. In CMS, a classifier trained on
expert-labelled data of 2010 was proposed to certify the clear cases automatically \cite{Borisyak:2017CMSDQ};
convolutional networks and autoencoders were trained on occupancy maps of the muon drift tubes
\cite{Pol:2018DT}, and autoencoders on summary statistics of reconstructed-object distributions in certified
luminosity sections of 2016 \cite{Pol:2019CMSDQM}; autoencoders \cite{CMSECAL:2024CSBS,Harilal:2025CALOR}
and graph networks \cite{Asres:2023HCAL} have been developed for the online monitoring of the
electromagnetic and hadron calorimeters; and AutoDQM, which combines a statistical test,
principal-component analysis and autoencoders on the histograms of the muon systems and the level-1
trigger, was evaluated on the runs of 2022 with good and bad labels \cite{Brinkerhoff:2026AutoDQM}. Kernel
methods have been used as a goodness-of-fit test for drift-tube chambers recording cosmic muons
\cite{Grosso:2023ltd}. DINAMO \cite{Gavrikov:2025DINAMO} follows evolving histogram templates with an
exponentially weighted moving average and a transformer-encoder variant, tested on synthetic data and
commissioned at LHCb, and MEDIC \cite{Bassa:2026MEDIC} detects and localises calorimeter regions switched
off in a fast detector simulation. These studies use either data internal to an experiment, which cannot be
shared, or synthetic and simulated histograms, which lack the run-to-run variation of real data. As a
consequence no two methods have been compared on the same real data with known ground truth, and
evaluations rarely fix the false-alarm rate on data recorded after the calibration data, which is the only
operating condition that matters in practice. In the search for new physics, public benchmark data sets
\cite{Kasieczka:2021LHCO,Aarrestad:2022DarkMachines,Govorkova:2022dataset} have made anomaly-detection
methods comparable; to our knowledge, data-quality monitoring has no such benchmark built on real collision
data. \textit{The obstacle is the data, not
the methods.}

The ATLAS Open Data release for research \cite{ATLAS:OpenDataPHYSLITE,Albornoz:2025ICHEP} removes this
obstacle. It contains the
proton--proton collision data recorded by ATLAS \cite{ATLAS:detector} in 2015 and 2016 at
$\sqrt{s}=13$~TeV, more than $7\times10^{9}$ events in the PHYSLITE format \cite{ATLAS:Run3Software},
released under the CC0 licence, and the files are not filtered by the GRLs. Every event carries its run and
LB number and the detector error flags of the event header, so LB-level monitoring histograms of
reconstructed objects can be rebuilt by anyone, with the GRLs as a real-data label.

In this paper we turn this release into a benchmark for LB-level DQ monitoring and use it to compare
detectors under a deployment-like protocol. The benchmark (Sect.~\ref{sec:data}) consists of 1289-bin
histograms for 39\,828 LBs of 217 runs, eight types of physics-motivated detector defects injected at
event level in the same events, so that the clean and the defective histograms are known for every LB, and
two real-data tracks. The protocol (Sect.~\ref{sec:protocol}) trains and calibrates every detector on
earlier runs only, fixes the false-positive rate at 1\% by split-conformal calibration, and compares the
efficiency of blind detectors with that of a test that knows the defect pattern, under Poisson noise and on
the real residuals. On this benchmark (Sects.~\ref{sec:methods} and
\ref{sec:results}) we show that references built across runs fail, that residuals with respect to each
LB's own run are the representation in which defects become visible, and that a learned classifier
on these residuals finds 43\% of the LBs with injected defects at 1\% false-positive rate, against 25\% for a
matched-filter scan. Part of this gain comes from knowing the defect catalogue; for defect types absent
from training, the classifier stays above the scan only when its training includes pattern-agnostic
synthetic detector failures.

The three findings are \textit{one} property of the data seen from three sides. The run-to-run
differences of trigger menus, pile-up and detector conditions that make references built across runs fail
are largely absent within a run, where the trigger menu and the detector configuration are fixed and the
pile-up decreases smoothly during the fill, which is why within-run residuals are close to
Poisson-distributed; and because
the residual null distribution is then nearly the same in every run, a classifier trained on the residuals
of earlier runs transfers to later ones. What limits all detectors at a 1\% false-alarm rate is what
remains: benign real-data anomalies inside certified runs, and defects whose signature is weaker than the
statistical precision of one LB.

The paper is organised as follows. Section~\ref{sec:data} describes the data, the histograms and the
injected defects, Sect.~\ref{sec:protocol} the evaluation protocol and the known-pattern bounds,
Sect.~\ref{sec:methods} the detectors, Sect.~\ref{sec:results} the results on injected defects,
Sect.~\ref{sec:real} the real-data tracks, and Sect.~\ref{sec:discussion} the limitations before we
conclude in Sect.~\ref{sec:conclusions}.

\section{Benchmark construction}
\label{sec:data}

\subsection{Data and LB sampling}
\label{sec:sampling}

The release contains 317 runs (131 from 2015 and 186 from 2016) in 55\,620 PHYSLITE files, of which 217
runs appear in the 2015 and 2016 physics GRLs. Reading every event of every LB is neither necessary
nor affordable for histograms whose purpose is to be compared at fixed statistics, so for each of the 217
runs we select a contiguous window of up to 100 files, determine the LBs contained in each file, and
keep a block of 240 consecutive GRL-certified LBs (all of them for shorter runs), together with the
uncertified LBs inside the block. Within each LB we read the first $10^4$ events in file
order; the histograms of the first $10^3$ and $3\times10^3$ events of the same LBs are produced in
the same pass, so the three statistics levels are nested samples of the same events. The data are read
with uproot \cite{uproot} directly from the open-data storage. In total $3.87\times10^{8}$ events from
14\,864 files enter the benchmark, about 5\% of the release, in 39\,828 LBs (38\,352 in the GRL)
with a median of 9\,923 events per LB.

Events are cleaned with the standard ATLAS event-cleaning criteria: events with the error state set for
the liquid-argon (LAr) calorimeter, the Tile calorimeter or the semiconductor tracker (SCT), incomplete
events, and events that fail the \textit{LooseBad} jet-cleaning requirement \cite{ATLAS:JetCleaning} are
rejected. This removes 1.1\% of the events, apart from one run discussed in Sect.~\ref{sec:runlevel}. The uncleaned histograms are stored for
the top statistics level, so that the effect of the cleaning can be studied.

\subsection{Monitored histograms}
\label{sec:hists}

For every LB we fill 20 monitored histograms with 1259 bins in total from the calibrated analysis objects
stored in PHYSLITE, without identification, isolation or jet-vertex-tagger requirements beyond the PHYSLITE
preselection: anti-$k_t$ jets with radius parameter $R=0.4$ built from particle-flow objects
\cite{Cacciari:2008gp,ATLAS:PFlow} with $p_\mathrm{T}>25$~GeV and $|\eta|<4.5$, electrons with
$p_\mathrm{T}>15$~GeV and $|\eta|<2.47$, photons with $p_\mathrm{T}>20$~GeV and $|\eta|<2.37$, and muons
with $p_\mathrm{T}>15$~GeV and $|\eta|<2.7$. For each object type there are the $\eta$, $\phi$ and
$\log p_\mathrm{T}$ distributions, for jets and muons an $\eta\times\phi$ occupancy map, for jets, electrons
and photons the $p_\mathrm{T}$ spectra in $\eta$ slices, and the jet multiplicity; the magnitude and azimuth
of the negative vectorial sum of the transverse momenta of the selected objects, $\vec{H}_\mathrm{T}^\mathrm{miss}$,
complete the set. The number of reconstructed primary vertices (30 bins) is stored but not monitored, which
brings the total to 1289 bins. The mean number of interactions per bunch crossing $\langle\mu\rangle$ of
each LB is stored as the condition variable. The histograms resemble those of the physics-object DQ
monitoring; the detector-level histograms of the subdetector monitoring (cell energies and timing, hit
maps, readout errors) cannot be built from PHYSLITE.

\subsection{Injected detector defects}
\label{sec:defects}

Real defects in the release are labelled only through the GRLs, and the reasons for rejecting an LB are not
public. To obtain a ground truth, we inject defects at event level into the selected objects of the same
events and fill for every LB both the clean and the injected histograms. The catalogue contains eight
types, each at three strengths:
\begin{itemize}
\item \textit{dead calorimeter region}: jets, electrons and photons in a $0.4\times0.4$ region in
$(\eta,\phi)$ are lost with probability 0.3, 0.6 or 0.95, and the surviving jets lose 30\% of their
$p_\mathrm{T}$ (a dead LAr front-end board or a LAr high-voltage sector switched off);
\item \textit{Tile module off}: jets within $|\Delta\phi|<0.15$ of one of the 64 modules of one Tile
partition (LBA, LBC, EBA or EBC) keep 90\%, 80\% or 65\% of their $p_\mathrm{T}$;
\item \textit{hot spot}: a fraction 0.3\%, 1\% or 4\% of the events contains a fake jet or photon at a
fixed $(\eta,\phi)$ (a noisy calorimeter cell);
\item \textit{noise burst}: a fraction 0.2\%, 0.8\% or 3\% of the events contains several fake jets and
photons in one LAr end-cap (coherent noise);
\item \textit{e/$\gamma$ scale shift} and \textit{jet scale shift}: the energy scale changes by 1, 3 or 8\%
(e/$\gamma$) and by 3, 6 or 12\% (jets) in one $\eta$ region, upwards or downwards (a miscalibration);
\item \textit{muon sector off}: muons in one of the 16 $\phi$ sectors of the muon spectrometer, on one or
both sides, are lost with probability 0.3, 0.6 or 0.95;
\item \textit{partition dropout}: a fraction 3\%, 10\% or 30\% of the events loses all calorimeter
objects in the $\eta$ range of one partition of the LAr electromagnetic calorimeter (a busy or
desynchronised readout partition).
\end{itemize}
Defects are placed in windows of 2--12 consecutive certified LBs. Windows are proposed with a Poisson mean
of five per 100 LBs, and a proposal is discarded if it overlaps or touches an existing window, so that
windows are separated by at least one clean LB; type, strength, position and direction are drawn at
random. All random numbers are derived from a hash of the event number and the window
identifier, so the injection does not depend on how files are split. In total 1359 windows cover 9\,226
LBs, 24\% of the certified ones.

\subsection{Real-data labels}
\label{sec:reallabels}

Two labels come from the real data. The 1\,476 uncertified LBs inside the sampled blocks are the
positives of a weak-label track: they lie at the start or the end of a run, have no recorded luminosity, or
lie inside the run and were rejected for reasons that are not public. The second track uses
the LAr error state in the event header, which is set for events inside the time-window veto around LAr
noise bursts and for events with corrupted LAr data \cite{ATLAS:LArDQ}: certified LBs in which at least 1\%
of the events are flagged are positives, those with less than 0.1\% are negatives, and the uncleaned
histograms are scored.

\section{Evaluation protocol}
\label{sec:protocol}

\subsection{Time-ordered evaluation}
\label{sec:split}

A detector in operation is trained and calibrated on data recorded before the data it monitors. We reproduce
this in a rolling split: within each year the runs are ordered in time and cut into five blocks; block
$i\ge2$ is the test set of fold $i$, and all earlier runs of the same year form its reference set, whose
latest 30\% serve for calibration when a method needs a separate calibration set. This gives four folds per
year, eight in total, with 13 (2015) or 30 (2016) test runs and 14--120 reference runs per fold (54 on
average); 172 runs with 23\,328 clean and 7\,293 injected certified LBs are tested. Folds never mix
the two years, whose pile-up profiles and trigger menus differ. As a robustness test we also train on 2015
and test on 2016 (\textit{year split}).

\subsection{Calibration and metrics}
\label{sec:metrics}

Every detector produces a score per LB. Its alarm threshold at false-positive rate $\alpha=1\%$ is the
split-conformal quantile of the scores of clean reference LBs, the $\lceil (n+1)(1-\alpha)\rceil$-th
smallest of $n$ \cite{Vovk:2005ALRW,Angelopoulos:2021conformal}, set separately in every fold. If the
reference and test LBs were exchangeable, this would guarantee an expected false-positive rate of at most
$\alpha$; for real data taken weeks apart it is an assumption, which we test by measuring the
false-positive rate on the clean test LBs. We report this \textit{conformal} false-positive rate and the true-positive rate (detection
efficiency) on injected test LBs. Because two detectors with different conformal false-positive rates
cannot be ranked by their efficiencies, we also report the efficiency at a threshold set to the 99\%
quantile of the clean test LBs themselves (\textit{TPR at test FPR 1\%}), which compares all
detectors at exactly 1\% false-positive rate. Scores of different folds are pooled after a tail
normalisation (median of the fold's clean reference scores $\to0$, their 99\% quantile $\to1$).
LBs of one run are correlated, so statistical intervals are obtained by resampling test runs with
replacement (1000 bootstrap samples); differences between detectors are evaluated on the same samples.

\subsection{Known-pattern bounds}
\label{sec:bounds}

Since the clean and injected histograms of each LB are known, the detectability of every injected
defect can be quantified independently of any detector. For an LB with clean counts $c_b$ and
injected counts $j_b$ in the most affected histogram, the significance of a test that knows the defect
pattern under Poisson fluctuations is
\begin{equation}
Z = \Big[\sum_b \frac{(j_b-c_b)^2}{c_b+1}\Big]^{1/2},
\label{eq:Z}
\end{equation}
where the $+1$ regularises empty bins. Such a test detects the defect at false-positive rate $\alpha$
with probability $\Phi(Z-z_\alpha)$, where $\Phi$ is the cumulative standard normal distribution and
$z_{0.01}=2.33$. We call the average of this probability over the injected LBs the \textit{IDEAL}
efficiency. The second reference, \textit{KNOWN(real)}, applies the true pattern of the most affected
histogram as a matched filter to the real residuals of Sect.~\ref{sec:runref}, with its threshold at the
99\% quantile of the absolute filter output over the clean test LBs: it includes the real noise and the uncertainty of
the real reference, but still knows the pattern. Both are bounds for tests that use a single histogram: a
detector that combines several histograms affected by the same defect could in principle exceed them, but
none of the detectors studied here does at the same false-positive rate, overall or in any bin of $Z$ or
defect type. The step
IDEAL\,$\to$\,KNOWN(real) measures the cost of real data, the step KNOWN(real)\,$\to$\,blind detector the
cost of not knowing what to look for.

\section{Detectors}
\label{sec:methods}

\subsection{References across runs}
\label{sec:baselines}

The baselines follow the designs used in DQ systems and in the literature. The \textit{static reference}
compares the bin fractions and object rates of every histogram with their mean over the clean reference
LBs by a $\chi^2$ whose variance includes the observed spread beyond the Poisson expectation; the
$\mu$-\textit{binned} and $\mu$-\textit{polynomial} references let the reference depend on
$\langle\mu\rangle$ in five quantile bins or quadratically. The \textit{EWMA} reference is updated LB
by LB with an exponentially weighted moving average (weight 0.1 for the newest LB), using every
LB known to be clean as if a shifter had confirmed it, similar to the statistical variant of DINAMO
\cite{Gavrikov:2025DINAMO}. The unsupervised learners are principal-component reconstruction, an
autoencoder, an autoencoder conditioned on $\langle\mu\rangle$, and an isolation forest
\cite{Liu:2008iforest}, all trained on the standardised bin fractions and rates of the clean histograms of
up to 6000 LBs of the training runs and implemented with scikit-learn \cite{Pedregosa:2011sklearn}.

\subsection{Within-run reference and scan}
\label{sec:runref}

Offline DQ assessment has the whole run available, and within a run most of the conditions that make runs
differ are fixed: the trigger menu and the detector configuration do not change, and the pile-up decreases
smoothly with the instantaneous luminosity of the fill. Trigger prescales, in contrast, can be changed
during a run \cite{ATLAS:TriggerRun2}, which changes the event composition of the stream in steps that the
reference described below does not model. For every
run and histogram we therefore fit the bin fractions of all LBs linearly in
$\langle\mu\rangle$ by iteratively re-weighted least squares with Huber weights \cite{Huber:1964} per
LB and bin, so that a defect touching many LBs in one bin, or many bins in a few LBs,
cannot pull the reference. An extra variance per bin is estimated from the median absolute deviation of the
residuals beyond the Poisson expectation, and every LB is compared with the reference fitted without
it: the leave-one-out residual divided by its full standard deviation, which includes the uncertainty of the
reference, is the \textit{pull} of each bin, approximately standard normal for clean LBs. The same
model is fitted to the object rates. The \textit{within-run reference} detector scores an LB by the
largest of the per-histogram mean and maximum squared pulls, each converted into a one-sided robust $z$ on
the reference LBs.

The \textit{within-run scan} adds matched filters for the patterns detector failures produce: for each
histogram the largest squared sum of pulls over contiguous windows of 1--16 bins ($\phi$ periodic) and over
$\eta\times\phi$ or $\eta\times p_\mathrm{T}$ rectangles, normalised to unit variance, and filters for a
shift or a change of width of ordered distributions, built from the first and second derivative of the
reference. The score is the maximum over all filters of their robust $z$; it is a scan statistic with a
trials factor that the conformal calibration absorbs.

\subsection{Learned residual detector}
\label{sec:learned}

The maximum over filters is powerful when a single filter captures the signal. Real defects, however,
spread over several histograms and bins in characteristic ways, and a classifier trained on examples can
learn which combinations matter. We compute, for every LB, 690 position-invariant
features of its pulls: per histogram the mean squared pull, the largest positive and negative window sums
at several widths, the largest rectangle sums of the two-dimensional maps, shift and width filters
(per $\eta$ slice for the $p_\mathrm{T}$ spectra), the rate pulls, the distance to the first and the last
LB of the sampled block, and global maxima over histograms. The features do not encode \textit{where} a deviation
sits, only its size and shape, so a model trained on a defect at one position applies to all positions.

The classifier is a histogram-based gradient-boosted decision-tree ensemble
\cite{Friedman:2001gbm,Ke:2017LightGBM} as implemented in scikit-learn \cite{Pedregosa:2011sklearn} (300
iterations at most, learning rate 0.08, 31
leaves, $L_2$ regularisation 1, early stopping on 15\% of the training data). It is trained on the
reference runs of a fold only. Negatives are the clean certified LBs; positives are the injected
LBs of the reference runs and, three per clean LB, synthetic defects: the fraction change of a
randomly chosen injected LB, rotated in $\phi$ by a multiple of $2\pi/16$ or mirrored in $\eta$ and
scaled by a log-uniform factor between 0.3 and 2, is added in pull space to a clean LB with that
LB's own uncertainties. Cross-fitting over five groups of reference runs provides out-of-sample
scores of all clean reference LBs for the conformal threshold. We average three models with
different random seeds.

Detector problems usually persist over several LBs, whereas many benign anomalies of real data, such as
those of the last LB of a run, affect a single LB. The configuration with temporal context (\Ct) adds the maximum and the minimum
over the two neighbouring LBs of 64 summary features (128 features in total), and draws its
synthetic defects in windows of 2--12 consecutive clean LBs; the clean LBs next to each
synthetic window are added as negatives, because otherwise the model learns that a defective neighbour
implies a defect. The configuration without context is denoted \Cs.

Finally, a detector trained on injected defects may miss defects unlike any in training. The
\textit{safety net} combines the learned score with the scan score by the maximum of the two tail-normalised
scores, so that an LB is flagged if either detector finds it unusual.

Two further options target defects outside the catalogue (Sect.~\ref{sec:unseen}). With \textit{monotonic
constraints} the score cannot decrease when any deviation-magnitude feature (window and rectangle maxima,
mean squared pulls, global maxima) grows, so that a deviation larger than any seen in training cannot be
scored as benign. The \textit{generic detector-failure prior} replaces half of the synthetic defects by
patterns that are not derived from the injected ones: an excess or deficit in a window or rectangle of one
histogram (30\%); objects lost or added in an $\eta\times\phi$ region of random position and size,
consistently in all histograms of one object collection, of the calorimeter objects or of all objects,
including the jet multiplicity (50\%); and an energy-scale shift of one collection in an $\eta$ range
(20\%). Deficits are drawn with probability 0.6, and the amplitude follows from a target known-signal
significance drawn log-uniformly between 2 and 80, capped at the physical maximum (for a deficit, all
affected objects lost). We denote the configurations by \Ct+m, \Ct+g and
\Ct+g+m, and call \textit{generic only} the variant of \Ct+g that draws all its synthetic defects from the
prior and uses no injected LB in training at all.

\begin{table*}[t]
\centering
\caption{Rolling split, $10^4$ events per LB (172 test runs, 23\,328 clean and 7\,293 injected
LBs). Conformal: false-positive and true-positive rate with the threshold calibrated on the reference
runs of each fold. Test FPR 1\%: efficiency at the 99\% quantile of the clean test LBs; the interval
is the 68\% range of a bootstrap over test runs, and the last column the difference to the within-run scan
with its 95\% interval from the same bootstrap samples. KNOWN(real) and IDEAL are the single-histogram
known-pattern bounds of Sect.~\ref{sec:bounds}.}
\label{tab:main}
\footnotesize
\setlength{\tabcolsep}{4pt}
\begin{tabular}{lcccc}
\toprule
Detector & Conformal FPR & Conformal TPR & TPR at test FPR 1\% & $-$ within-run scan \\
\midrule
Learned, temporal context (\Ct)      & 0.010 & 0.433 & 0.433 [0.398, 0.456] & $+0.186$ [$+0.125$, $+0.215$] \\
\Ct\ + scan safety net               & 0.011 & 0.402 & 0.393 [0.360, 0.427] & $+0.145$ [$+0.104$, $+0.184$] \\
Learned, single LB (\Cs)      & 0.008 & 0.358 & 0.388 [0.353, 0.420] & $+0.141$ [$+0.091$, $+0.178$] \\
Within-run scan                      & 0.010 & 0.245 & 0.247 [0.228, 0.267] & --- \\
Within-run reference                 & 0.009 & 0.242 & 0.249 [0.235, 0.265] & $+0.002$ [$-0.016$, $+0.030$] \\
EWMA reference                       & 0.013 & 0.177 & 0.165 [0.151, 0.178] & $-0.082$ [$-0.115$, $-0.051$] \\
Static reference                     & 0.091 & 0.192 & 0.069 [0.062, 0.082] & $-0.178$ [$-0.213$, $-0.136$] \\
$\mu$-binned reference               & 0.194 & 0.270 & 0.054 [0.048, 0.066] & $-0.193$ [$-0.230$, $-0.150$] \\
$\mu$-polynomial reference           & 0.142 & 0.248 & 0.063 [0.056, 0.079] & $-0.184$ [$-0.219$, $-0.136$] \\
PCA reconstruction                   & 0.011 & 0.020 & 0.029 [0.024, 0.036] & $-0.218$ [$-0.255$, $-0.175$] \\
Autoencoder                          & 0.011 & 0.018 & 0.029 [0.023, 0.035] & $-0.218$ [$-0.256$, $-0.176$] \\
Conditional autoencoder              & 0.014 & 0.023 & 0.029 [0.023, 0.035] & $-0.218$ [$-0.255$, $-0.175$] \\
Isolation forest                     & 0.019 & 0.022 & 0.013 [0.010, 0.015] & $-0.234$ [$-0.270$, $-0.194$] \\
\midrule
KNOWN(real) (known pattern)          & 0.010 & \multicolumn{2}{c}{0.626} & \\
IDEAL (known pattern, Poisson)       & 0.010 & \multicolumn{2}{c}{0.765} & \\
\bottomrule
\end{tabular}
\end{table*}

\section{Results on injected defects}
\label{sec:results}

\subsection{Run-to-run variation}
\label{sec:variability}

\begin{figure}[t]
\centering
\includegraphics[width=\columnwidth]{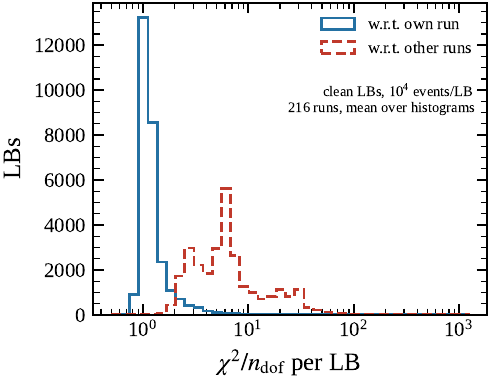}
\caption{$\chi^2/n_\mathrm{dof}$ of clean certified LBs ($10^4$ events per LB), averaged over the
one-dimensional histograms and the $\eta\times\phi$ maps, with respect to the median of their own run
(solid) and to the median of the medians of all other runs (dashed). LBs with fewer than 80\% of the median
number of events are not included; run 300908, which has no events left after the event cleaning, does not
enter (216 runs). Within a run the shapes agree to Poisson precision; across runs they do not.}
\label{fig:variability}
\end{figure}

\begin{figure*}[t]
\centering
\includegraphics[width=0.49\textwidth]{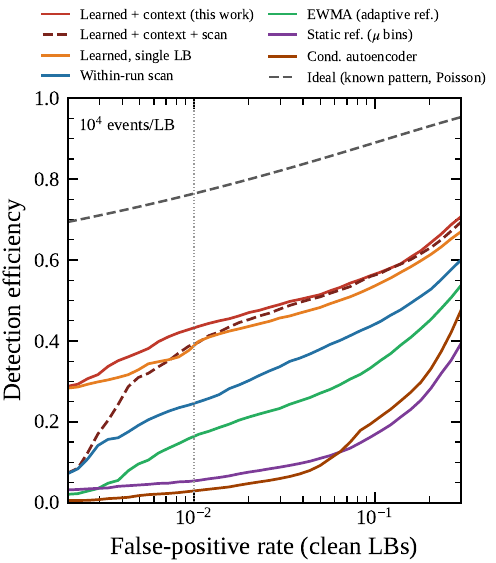}\hfill
\includegraphics[width=0.49\textwidth]{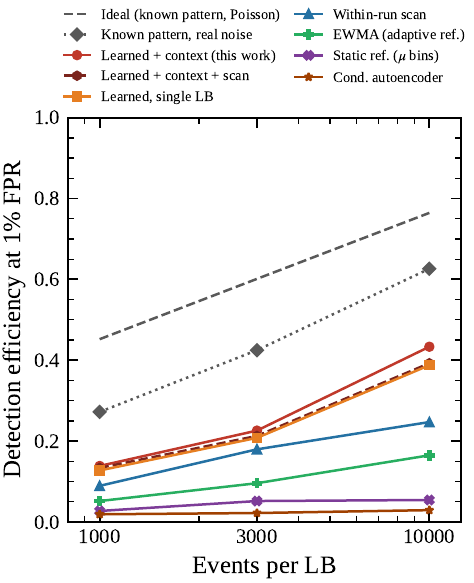}
\caption{Left: detection efficiency as a function of the false-positive rate on clean test LBs,
rolling split, $10^4$ events per LB. Right: efficiency at 1\% false-positive rate as a function of
the number of events per LB (nested samples of the same events), with the two known-pattern bounds of
Sect.~\ref{sec:bounds}.}
\label{fig:roc}
\end{figure*}

Figure~\ref{fig:variability} shows why references built across runs cannot work at the statistical
precision of an LB: compared with the median of its own run, the median clean LB has
$\chi^2/n_\mathrm{dof}=1.11$, close to the Poisson expectation, while compared with the other runs it has
5.90. The differences between runs (trigger menus, pile-up profiles, detector conditions) are thus several
times larger than the statistical fluctuations, and they are coherent over many bins, which is exactly the
signature of a defect.

\subsection{Detection efficiency at 1\% false-positive rate}
\label{sec:main}

\begin{figure*}[t]
\centering
\includegraphics[width=0.49\textwidth]{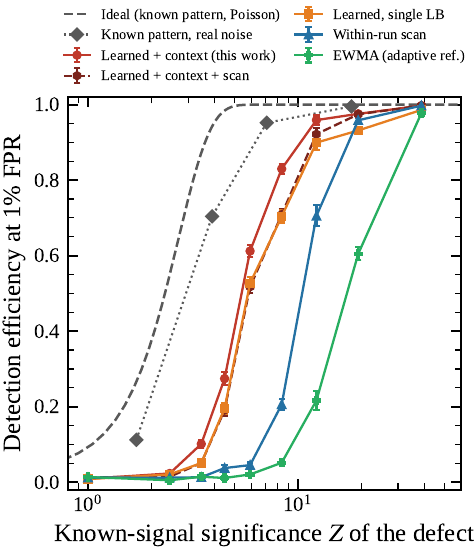}\hfill
\includegraphics[width=0.49\textwidth]{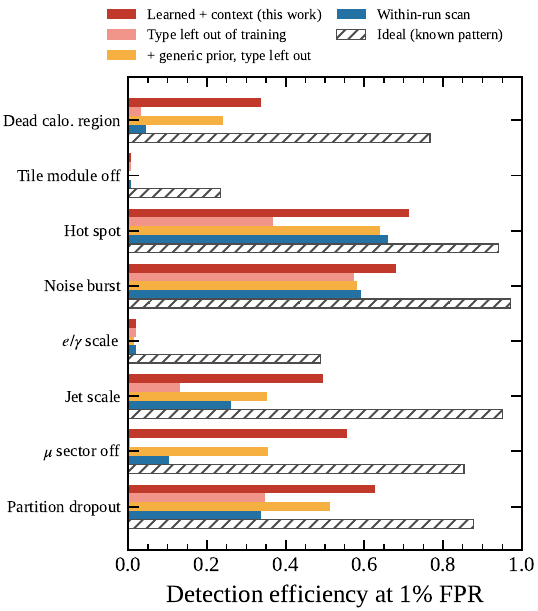}
\caption{Left: efficiency at 1\% false-positive rate as a function of the known-signal significance $Z$ of
Eq.~(\ref{eq:Z}), rolling split, $10^4$ events per LB; the dashed line is $\Phi(Z-2.33)$. Right:
efficiency per defect type (all strengths) for the learned detector, the same detector when the type was
left out of training, without and with the generic detector-failure prior of Sect.~\ref{sec:learned}
(\Ct+g; learned detector alone, one seed), the within-run scan, and the IDEAL bound.}
\label{fig:loss}
\end{figure*}

Table~\ref{tab:main} and Fig.~\ref{fig:roc} summarise the comparison. As expected from
Fig.~\ref{fig:variability}, every detector that compares an LB with other runs fails at a 1\%
false-alarm budget, in one of two ways: the explicit references cannot reach the target false-positive rate
at all (9--19\% conformal false-positive rate, and 5--7\% efficiency when forced to 1\%), while the
unsupervised learners keep the false-positive rate but detect 1--3\%, because their reconstruction error is
dominated by run-to-run differences that they cannot model from a few thousand training LBs. The
EWMA reference, which follows the conditions in time, reaches 17\%.

The within-run reference and the scan reach 25\%, with conformal false-positive rates of 0.9--1.0\%: the
residual representation alone does most of the work, and the matched filters of the scan add little on top
of the maximum pull at this statistics. The learned detector on the same residuals reaches 36\% for single
LBs and 43\% with temporal context, at conformal false-positive rates of 0.8\% and 1.0\%; the gain
over the scan, $+0.19$ with a 95\% interval of $[+0.13, +0.22]$, is robust against the run-to-run
correlation of the test LBs. The three seeds give 0.432, 0.424 and 0.433 at test FPR 1\%, so the
model variance is small compared with the run-to-run uncertainty. The safety net costs 0.04 for the defect
types in the catalogue; its purpose is the robustness to other types studied in
Sect.~\ref{sec:unseen}.

The known-pattern bounds put these numbers in perspective. A test that knows the defect pattern would detect
77\% under Poisson fluctuations and 63\% on the real residuals of the most affected histogram: one fifth of the detectable defects is lost to
real-data noise and reference uncertainty before any search for an unknown pattern starts, and the learned
detector reaches 69\% of the KNOWN(real) efficiency. The same ordering holds at lower statistics
(Fig.~\ref{fig:roc}, right): with $3\times10^3$ events per LB the learned detector reaches 23\%
against 18\% for the scan (KNOWN(real) 43\%), and with $10^3$ events 14\% against 9\% (27\%).

\subsection{Where the efficiency is lost}
\label{sec:loss}

Figure~\ref{fig:loss} (left) resolves the efficiency by the known-signal significance $Z$. Defects with
$Z>10$ are found by all residual detectors (98\% learned, 93\% scan), and defects with $Z<3$ by none
(2\% and 1\%, against 35\% for the IDEAL test), as anticipated from the statistics of a single LB. The
learned detector gains in between: for $5<Z<10$ it detects 71\% against 12\% for the scan and 95\% for
KNOWN(real), and for $3<Z<5$ 19\% against 2\% (KNOWN(real) 70\%). This is where combining many weak
deviations matters, because a defect of moderate significance distributes it over several bins and
histograms, none of which crosses the scan threshold alone.

By type (Fig.~\ref{fig:loss}, right; conformal efficiency at strengths 1/\allowbreak 2/\allowbreak 3), the learned detector finds
hot spots (0.12/\allowbreak 0.92/\allowbreak 1.00), noise bursts (0.35/\allowbreak 0.92/\allowbreak 1.00), partition dropouts (0.02/\allowbreak 0.82/\allowbreak 1.00), muon
sectors (0.05/\allowbreak 0.55/\allowbreak 0.95), jet-scale shifts (0.05/\allowbreak 0.38/\allowbreak 0.91) and dead calorimeter regions (0.04/\allowbreak 0.20/\allowbreak 0.80);
the scan finds the strong instances of the first three but only 25\% of the strong muon-sector losses and
11\% of the strong dead regions. Two types remain nearly invisible to every detector: Tile modules switched
off ($\le1\%$) and e/$\gamma$ scale shifts ($\le6\%$). For Tile modules the bounds (IDEAL 0.08/\allowbreak 0.18/\allowbreak 0.42,
KNOWN(real) 0.02/\allowbreak 0.05/\allowbreak 0.14) indicate that the limitation is the information content of the object-level
histograms at $10^4$ events, not the detectors: only the jets within $|\Delta\phi|<0.15$ of one module in
one partition are affected, and a 10--35\% loss of their $p_\mathrm{T}$ moves few of them across bin
edges. The same holds for the two weaker e/$\gamma$ scale shifts (KNOWN(real) 0.03
and 0.10), but an 8\% shift would be found in 65\% of the cases by a test that knows its pattern: there the
blind detectors lose a smooth distortion of sparsely populated $p_\mathrm{T}$ spectra in the trials of
searching all histograms and regions, which makes this type a target for future detectors.

\subsection{Calibration over time}
\label{sec:calibration}

\begin{table*}[t]
\centering
\caption{Learned-detector configurations, rolling split, $10^4$ events per LB, efficiency at test
FPR 1\% (one seed; \Ct\ with three seeds: 0.433). Known types: every type in training. Unseen types: for
the LBs of each type a model trained without that type. Differences: paired bootstrap over test
runs, 95\% intervals. The generic-only detector uses no injected LB in training, so every type is
unseen for it. Thresholds are re-estimated in every bootstrap sample, except for
the fixed unseen-type decisions.}
\label{tab:unseen}
\footnotesize
\setlength{\tabcolsep}{3.5pt}
\begin{tabular}{lcccc}
\toprule
Configuration & Known types & $-$ \Ct & Unseen types & $-$ within-run scan \\
\midrule
\Ct\ (catalogue)                & 0.432 & ---                           & 0.179 & $-0.069$ [$-0.103$, $-0.032$] \\
\Ct\ + scan safety net          & 0.392 & $-0.040$ [$-0.052$, $-0.019$] & 0.225 & $-0.022$ [$-0.052$, $+0.010$] \\
\Ct+m                           & 0.411 & $-0.021$ [$-0.031$, $-0.008$] & 0.186 & $-0.061$ [$-0.094$, $-0.027$] \\
\Ct+g                           & 0.403 & $-0.029$ [$-0.093$, $-0.012$] & 0.338 & $+0.090$ [$+0.058$, $+0.127$] \\
\Ct+g + scan safety net         & 0.357 & $-0.075$ [$-0.110$, $-0.045$] & 0.271 & $+0.024$ [$-0.005$, $+0.058$] \\
\Ct+g+m                         & 0.374 & $-0.058$ [$-0.096$, $-0.034$] & 0.304 & $+0.057$ [$+0.025$, $+0.090$] \\
Generic only                    & ---   & ---                           & 0.357 & $+0.110$ [$-0.018$, $+0.146$] \\
Within-run scan                 & 0.247 & ---                           & 0.247 & --- \\
\bottomrule
\end{tabular}
\end{table*}

The split-conformal threshold assumes that the clean LBs of the test runs are distributed like those
of the reference runs. In the rolling split the pooled conformal false-positive rate of the learned detector
is 1.0\% at $10^4$ events (1.3\% at $10^3$ and $3\times10^3$), with six of eight folds between 0.1\% and
0.7\% and two folds at 2.3\% (fourth block of 2015, 40 reference runs) and 3.1\% (second block of 2016,
30 reference runs). The second of these is due to a single certified run, 301918: in the seed-0 model, 79
of its 140 clean test LBs exceed the conformal threshold, against 26 in the 29 other test runs of the
block, and without this run the fold's false-positive rate is 0.7\% instead of 2.6\%. The scan flags 40
LBs of the same run. The flagged LBs include every one of the last 43 clean LBs of the
run (from LB 170 on) and about half of those in two earlier stretches, which suggests that the conditions
change within the run, while its summed histograms are unremarkable at run level (robust $z=1.2$ in
Sect.~\ref{sec:runlevel}). A change of trigger prescales during the run would have this signature, but the
public data do not allow us to identify the cause. Training on 2015 and testing on 2016 violates the assumption more strongly: the conformal
false-positive rate of the learned detector rises to 2.5\% (\Ct) and 1.4\% (\Cs), and to 2.9--6.0\% at lower
statistics, while its efficiency at test FPR 1\% stays at 38\% against 24\% for the scan. The ranking of the
detectors is therefore stable under the changes of pile-up and trigger menu between the two years, but the
threshold is not, which
is why the reference must come from the same data-taking period.

\subsection{How much history is needed}
\label{sec:history}

\begin{figure}[t]
\centering
\includegraphics[width=\columnwidth]{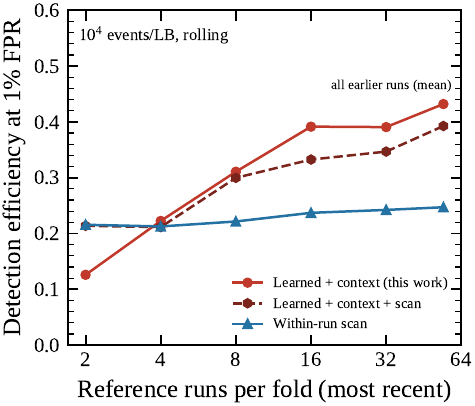}
\caption{Efficiency at 1\% false-positive rate (clean test LBs, rolling split, $10^4$ events per
LB) when every fold uses only its $N$ most recent reference runs for training, cross-fitting,
calibration and the scan normalisation; one seed. The right-most points use all earlier runs of the year
(54 on average).}
\label{fig:history}
\end{figure}

A detector that needs a long history is of little use at the start of a data-taking year. In
Fig.~\ref{fig:history} every fold uses only its $N$ most recent reference runs. With two runs the scan is
better (22\% against 13\%), and the conformal threshold of the learned detector is conservative (0.3\%
false-positive rate); with four runs the two are equivalent (22\% and 21\%); from eight runs on the learned
detector is clearly better (31\% against 22\%), it reaches 39\% with 16--32 runs and 43\% with all earlier
runs, while the scan stays between 21\% and 25\% for every $N$. From four runs on, the conformal false-positive rate
of the learned detector stays between 0.7\% and 1.1\%. The learned detector thus converts additional
history into efficiency, whereas the scan, whose only fitted quantities are normalisations, saturates
early; a deployment would use the scan for the first few runs of a year.

\subsection{Defect types not seen in training}
\label{sec:unseen}

The synthetic training defects are derived from the injected ones, so part of the gain may come from
knowing the catalogue. We test this by leaving one type out at a time: for the test LBs of type $t$
we use a model trained without any type-$t$ pattern. The answer is unfavourable for the learned detector
alone: its efficiency on unseen types is 17.4\% (17.9\% at test FPR 1\%), below the 24.5\% of the scan. The
loss is large for dead calorimeter regions (0.04/0.01/0.04 at the three strengths, against 0.04/0.20/0.80
when the type is known) and muon sectors (0.00/0.01/0.00 against 0.05/0.55/0.95), and strong unseen defects
are sometimes scored as benign (hot spot strength 3: 0.54). A plausible reason is that deviations this large
occur in training only in the benign real-data anomalies, and decision trees do not extrapolate beyond the
range of their training data. With the safety net the efficiency on
unseen types is 22.8\%, and the strong instances of hot spots, noise bursts and partition dropouts are
recovered (1.00, 1.00 and 0.99).

Table~\ref{tab:unseen} compares the two options of Sect.~\ref{sec:learned} on the full data. Monotonic
constraints alone cost two percentage points on the known types and leave the unseen-type efficiency unchanged
(18.6\%); they do not prevent strong unseen hot spots from being scored as benign (0.53 at strength 3),
presumably because the model can still lower the score through features that are not constrained, such as
the temporal context. The generic detector-failure prior changes the picture. With half of the synthetic
defects drawn from it (\Ct+g), the efficiency on unseen types rises from 17.9\% to 33.8\% (paired
difference $+0.159$ [$+0.136$, $+0.181$]), nine percentage points above the scan and eleven above the
catalogue-trained detector with the safety net. When their type is unseen, dead calorimeter regions are
found in 24\% of the cases instead of 3\%, muon sectors in 35\% instead of 0\%, jet-scale shifts in 35\%
instead of 13\% and partition dropouts in 51\% instead of 35\% (Fig.~\ref{fig:loss}, right; the scan finds
5\%, 10\%, 26\% and 34\%). Hot spots and noise bursts, which the scan finds as well, are found about as
often as by the scan (64\% and 58\% against 66\% and 59\%). The prior also removes the failure that the
monotonic constraints were meant to fix: strong unseen hot spots, noise bursts and partition dropouts are
found in 99\%--100\% of the cases, and strong unseen muon-sector losses, jet-scale shifts and dead regions
in 75\%, 71\% and 58\%. With the prior, the monotonic constraints lower the efficiency on unseen types
(30.4\% for \Ct+g+m, paired difference to \Ct+g $-0.034$ [$-0.041$, $-0.026$]) and do not help on known
types (37.4\% against 40.3\%, $-0.029$ [$-0.037$, $+0.008$]). The safety net also lowers the efficiency
on unseen types once the prior is used (27.1\%), most likely because the scan finds few defects that the
prior-trained detector misses (Fig.~\ref{fig:loss}, right), while the two detectors share the 1\% false-alarm
budget.

The price of the prior is paid on the known types: 40.3\% instead of 43.2\%, lost at moderate significance
($3<Z<5$: 13\% against 19\%; $5<Z<10$: 65\% against 71\%), where the exact shape of a catalogued pattern
helps. A detector trained on the generic prior alone, which has never seen an injected LB, finds
35.7\% of all defects (34.1\% with monotonic constraints), about as much as \Ct+g on unseen types and 11
percentage points more than the scan. The 95\% interval of this difference, [$-0.018$, $+0.146$], reaches below zero
because of run 301918 (Sect.~\ref{sec:calibration}): it contains 87 of the 234 clean test LBs above
the threshold of this detector, and only bootstrap samples that contain it three times or more give a
negative difference; without this run the interval is [$+0.100$, $+0.150$].
Two caveats apply. The generic families reflect broad mechanisms of detector failure (local, regional, energy scale)
that also underlie our catalogue, so an unseen type here means an unseen pattern, not an unseen mechanism.
And the families and their proportions were fixed on a 48-run development sample of an early version of the
benchmark, which overlaps with the test runs; they were not tuned on the full data. We keep \Ct\ as the
reference result of the benchmark, because it measures what is attainable when the failure modes are known,
and recommend \Ct+g for operation, where the catalogue is never complete: it gives up three percentage
points on the known types for sixteen on the unknown ones and stays above the scan in both cases.

\section{Real data}
\label{sec:real}

\subsection{Uncertified LBs}
\label{sec:nongrl}

Of the 972 uncertified test LBs, 590 lie inside runs, 181 at the end and 47 at the start of runs,
and 154 have no luminosity. At the 1\% threshold the scan flags 22\% of them and the learned detector 16\%.
The lower value is expected: the learned detector was trained to recognise injected defect patterns and to
ignore the run-edge anomalies of certified data, and the reasons for rejecting most uncertified LBs
(beam conditions, a subdetector not yet ready for data taking, problems in subdetectors not represented in
our histograms) need not
produce any signature in reconstructed-object histograms. This track is therefore a weak label, useful to
show that the detectors respond to real conditions, not to rank them.

\subsection{LAr event veto}
\label{sec:larflag}

\begin{figure}[t]
\centering
\includegraphics[width=\columnwidth]{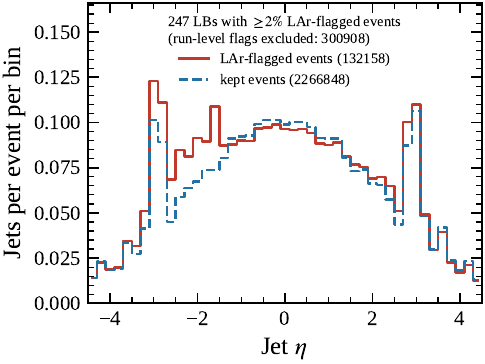}
\caption{Jets per event and $\eta$ bin in the certified LBs with at least 2\% LAr-flagged events,
for the flagged events (removed by the event cleaning) and the kept events; run 300908, in which every
event is flagged, is excluded.}
\label{fig:larflag}
\end{figure}

Excluding the run discussed below, 510 certified LBs have at least 1\% LAr-flagged events and 32\,248
less than 0.1\%. On the uncleaned histograms the within-run detectors separate the two classes with an area
under the ROC curve of 0.53--0.55, and the efficiency at 1\% false-positive rate is 6--7\%. The reason is
visible in Fig.~\ref{fig:larflag}: the flagged events contain more jets in the C-side end-cap
region ($-3.2\lesssim\eta\lesssim-1.4$) but are otherwise ordinary events, and they make up a few per cent of an LB,
so the LB histograms change by much less than their statistical precision. We report this as a
negative result: the event-level flag, not LB-level histograms of reconstructed objects, is the right
tool for noise bursts, which ATLAS removes with an event-level time-window veto \cite{ATLAS:LArDQ}.

\subsection{Run-level anomalies}
\label{sec:runlevel}

Within-run detectors are blind by construction to a problem that affects a whole run. In run 300908, which
is in the GRL, every one of the 2.35 million events read carries the LAr error state, so the event cleaning
removes the run entirely. Of the LAr-flagged events in certified LBs with at least 2\% flagged
events, 95\% come from this run, and these events contain almost no jets beyond $|\eta|\approx3.2$ and fewer than usual in
$2.5<|\eta|<3.2$, consistent with missing forward-calorimeter information.
Standard analyses that apply the event cleaning are not affected, but the example shows that a complete DQ
system needs a run-level comparison as well.

\begin{figure}[t]
\centering
\includegraphics[width=\columnwidth]{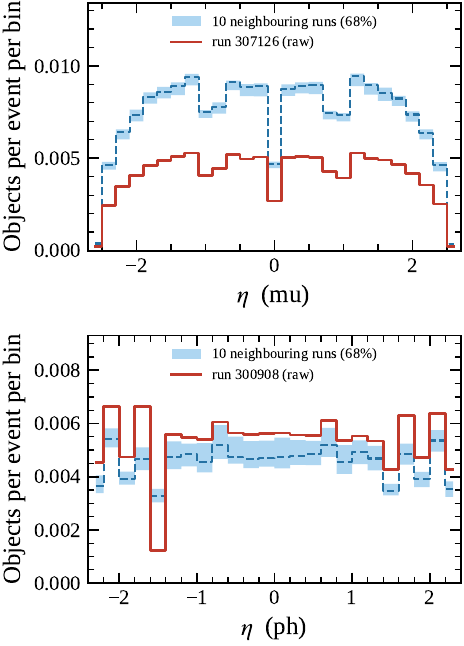}
\caption{Run-level comparison (uncleaned data): objects per event and bin of the most deviating
one-dimensional histogram for the highest-ranked run, 307126 (top, muon $\eta$), and for run 300908, in
which every event carries the LAr error state (bottom, photon $\eta$), with the 68\% band of the ten closest
runs of the same year.}
\label{fig:runlevel}
\end{figure}

We compare the summed histograms of every run with those of the ten closest runs of the same year, by the
shape $\chi^2/n_\mathrm{dof}$ and the rate pull of each histogram. Because the $p_\mathrm{T}$ spectra vary
much more between runs than the $\eta$ distributions, each histogram's deviation is divided by its median
over all runs, and the run score is the largest normalised deviation. In the uncleaned data this ranks run
300908 tenth of 215 (robust $z=2.3$ of the logarithmic score; eighth without the normalisation). Its
largest deviation is in the jet $\eta\times p_\mathrm{T}$ map, and its photon $\eta$ distribution has a
deficit near $\eta=-1.5$ and excesses at larger $|\eta|$ (Fig.~\ref{fig:runlevel}, bottom), but nine runs
deviate more. These runs are not detector problems that we can identify. In run 307126 ($z=3.0$) the muons per
event are about half of those in the neighbouring runs, with an unchanged $\eta$ distribution (rate pull
$-14.7$, Fig.~\ref{fig:runlevel}, top), and in run 284154 the photons per event are lower (rate pull
$-12.6$). A different trigger menu or prescale set would produce such differences, and the public data do
not allow us to separate it from a detector effect. A useful run-level detector
therefore needs references that model the trigger and beam conditions of each run, which we leave to future
work.

\section{Discussion and limitations}
\label{sec:discussion}

The benchmark contains reconstructed objects only. Subdetector DQ in the experiments works largely on
detector-level quantities (hit maps, cell energies and timing, readout errors), which are not in PHYSLITE,
so the defects we inject are those that reach the reconstructed objects; defects that do not, or that
reconstruction corrects, are outside its scope, and the Tile-module failures of Sect.~\ref{sec:loss} show
that some realistic defects remain invisible at this level.

The injected defects are simplified models of real failures. They are added to the objects of real events,
so the statistical fluctuations, the run-to-run variation and the benign anomalies of the data are real, but
their parameters were chosen by us, and the absolute efficiencies depend on that choice. The comparison
between detectors and with the known-pattern bounds of Sect.~\ref{sec:bounds} is made on the same
injected LBs and should depend on it less, but other defect models could change the ranking of
detectors that are close. The unseen-type
test of Sect.~\ref{sec:unseen} quantifies how much the learned detector depends on the catalogue.

The protocol reads at most $10^4$ events per LB and 240 LBs per run, about 5\% of the release.
The nested statistics levels show how the efficiencies scale with the number of events; for LBs with
more than $10^4$ recorded events, their full statistics would move all detectors and bounds upward. The run sample, in contrast, is complete: all 217 runs of the GRLs are used.

All computations run on one 16-core node of the CERN lxplus service. Indexing the release and building
the three nested datasets takes 6.8~h with 12 worker processes; the baselines and the bounds of one split
and statistics level take 1--6~min. The learned detector takes about 70~s per fold and seed with six
threads (28~min for the eight rolling folds and three seeds), and a leave-one-type-out evaluation, which
trains nine models per fold, about 3~h.

\section{Conclusions}
\label{sec:conclusions}

The decisive property of LB-level monitoring data is that most of the conditions which differ between
runs are fixed within a run. References built across runs therefore fail at a 1\% false-alarm budget, and
residuals with respect to a robust within-run reference make defects visible with nearly Poisson-distributed
noise. On these residuals a gradient-boosted classifier trained on synthetic defects and calibrated by
split-conformal prediction on earlier runs finds 43\% of the LBs with injected defects at a 1.0\% false-positive
rate, against 25\% for a matched-filter scan of the same residuals, and it overtakes the scan with as few as
eight earlier runs.

A test that knows the defect pattern finds 63\% of the injected defects on the real residuals of the most
affected histogram; the learned detector reaches 69\% of this, and about four fifths of the remaining gap
lie at known-signal significances between 3 and 10. Tile-module failures and the weaker electron/photon scale
shifts are below the information content of object-level histograms at $10^4$ events per LB, while
the strongest electron/photon scale shift is detectable in principle but missed by all blind detectors.
Trained on the defect catalogue alone, the learned detector falls below the scan for defect types it has
never seen (18\% against 25\%). Pattern-agnostic synthetic detector failures in its training raise this to
34\% at a cost of three percentage points on the catalogued types, and a detector trained on them alone finds 36\% of
all defects; monotonic constraints on the score do not help.

The benchmark, the injection code and the evaluation protocol are public, so that other detectors can be
compared on the same data under the same conditions. Natural extensions are the full statistics of every
LB, a run-level detector that complements the within-run methods, and defect models validated
against the defects recorded by the experiments.

\begin{acknowledgement}
\textbf{Acknowledgements} We acknowledge the work of the ATLAS Collaboration to record or simulate, reconstruct, and distribute the
Open Data used in this paper, and to develop and support the software with which it was analysed. The data
are released under the CC0 licence through the CERN Open Data Portal. Neither the ATLAS Collaboration nor
CERN endorses this work, and its results are not ATLAS results.

\end{acknowledgement}

\bibliographystyle{elsarticle-num}
\bibliography{refs}

@article{ATLAS:detector,
  author        = {{ATLAS Collaboration}},
  title         = {{The ATLAS Experiment at the CERN Large Hadron Collider}},
  journal       = {JINST},
  volume        = {3},
  pages         = {S08003},
  year          = {2008},
  doi           = {10.1088/1748-0221/3/08/S08003}
}

@article{ATLAS:DQ2015-2018,
  author        = {{ATLAS Collaboration}},
  title         = {{ATLAS data quality operations and performance for 2015--2018 data-taking}},
  journal       = {JINST},
  volume        = {15},
  pages         = {P04003},
  year          = {2020},
  doi           = {10.1088/1748-0221/15/04/P04003},
  eprint        = {1911.04632},
  archivePrefix = {arXiv}
}

@article{ATLAS:LArDQ,
  author        = {{ATLAS Collaboration}},
  title         = {{Monitoring and data quality assessment of the ATLAS liquid argon calorimeter}},
  journal       = {JINST},
  volume        = {9},
  pages         = {P07024},
  year          = {2014},
  doi           = {10.1088/1748-0221/9/07/P07024},
  eprint        = {1405.3768},
  archivePrefix = {arXiv}
}

@article{ATLAS:Run3Software,
  author        = {{ATLAS Collaboration}},
  title         = {{Software and computing for Run 3 of the ATLAS experiment at the LHC}},
  journal       = {Eur. Phys. J. C},
  volume        = {85},
  pages         = {234},
  year          = {2025},
  doi           = {10.1140/epjc/s10052-024-13701-w},
  eprint        = {2404.06335},
  archivePrefix = {arXiv}
}

@techreport{ATLAS:JetCleaning,
  author        = {{ATLAS Collaboration}},
  title         = {{Selection of jets produced in 13 TeV proton--proton collisions with the ATLAS detector}},
  institution   = {CERN},
  number        = {ATLAS-CONF-2015-029},
  year          = {2015}
}

@misc{ATLAS:OpenDataPHYSLITE,
  author        = {{ATLAS Collaboration}},
  title         = {{DAOD\_PHYSLITE format 2015--2016 Open Data for Research from the ATLAS experiment}},
  howpublished  = {CERN Open Data Portal},
  year          = {2024},
  doi           = {10.7483/OPENDATA.ATLAS.9HK7.P5SI}
}

@article{Albornoz:2025ICHEP,
    author = "Albornoz, Mariana Vivas",
    collaboration = "ATLAS",
    title = "{The First Release of ATLAS Open Data for Research}",
    doi = "10.22323/1.476.1172",
    journal = "PoS",
    volume = "ICHEP2024",
    pages = "1172",
    year = "2025"
}

@article{Borisyak:2017CMSDQ,
  author        = {Borisyak, Maxim and Ratnikov, Fedor and Derkach, Denis and Ustyuzhanin, Andrey},
  title         = {{Towards automation of data quality system for CERN CMS experiment}},
  journal       = {J. Phys. Conf. Ser.},
  volume        = {898},
  pages         = {092041},
  year          = {2017},
  doi           = {10.1088/1742-6596/898/9/092041},
  eprint        = {1709.08607},
  archivePrefix = {arXiv}
}

@article{Pol:2018DT,
  author        = {Pol, Adrian Alan and Cerminara, Gianluca and Germain, Cecile and Pierini, Maurizio and Seth, Agrima},
  title         = {{Detector monitoring with artificial neural networks at the CMS experiment at the CERN Large Hadron Collider}},
  journal       = {Comput. Softw. Big Sci.},
  volume        = {3},
  pages         = {3},
  year          = {2019},
  doi           = {10.1007/s41781-018-0020-1},
  eprint        = {1808.00911},
  archivePrefix = {arXiv}
}

@article{Pol:2019CMSDQM,
  author        = {Pol, Adrian Alan and Azzolini, Virginia and Cerminara, Gianluca and De Guio, Federico and Franzoni, Giovanni and Pierini, Maurizio and Siroký, Filip and Vlimant, Jean-Roch},
  title         = {{Anomaly detection using Deep Autoencoders for the assessment of the quality of the data acquired by the CMS experiment}},
  journal       = {EPJ Web Conf.},
  volume        = {214},
  pages         = {06008},
  year          = {2019},
  doi           = {10.1051/epjconf/201921406008}
}

@article{CMSECAL:2024CSBS,
  author        = {{CMS ECAL Collaboration}},
  title         = {{Autoencoder-based Anomaly Detection System for Online Data Quality Monitoring of the CMS Electromagnetic Calorimeter}},
  journal       = {Comput. Softw. Big Sci.},
  volume        = {8},
  pages         = {11},
  year          = {2024},
  doi           = {10.1007/s41781-024-00118-z},
  eprint        = {2309.10157},
  archivePrefix = {arXiv}
}

@article{Harilal:2025CALOR,
  author        = {Harilal, Abhirami and Park, Kyungmin and Paulini, Manfred},
  title         = {{Anomaly Detection Based on Machine Learning for the CMS Electromagnetic Calorimeter Online Data Quality Monitoring}},
  journal       = {EPJ Web Conf.},
  volume        = {320},
  pages         = {00048},
  year          = {2025},
  doi           = {10.1051/epjconf/202532000048},
  eprint        = {2407.20278},
  archivePrefix = {arXiv},
  note          = {On behalf of the CMS Collaboration}
}

@article{Asres:2023HCAL,
  author        = {Asres, Mulugeta Weldezgina and Omlin, Christian Walter and Wang, Long and Yu, David and Parygin, Pavel and Dittmann, Jay and Karapostoli, Georgia and Seidel, Markus and Venditti, Rosamaria and Lambrecht, Luka and Usai, Emanuele and Ahmad, Muhammad and Fernandez Menendez, Javier and Maeshima, Kaori and {CMS-HCAL Collaboration}},
  title         = {{Spatio-Temporal Anomaly Detection with Graph Networks for Data Quality Monitoring of the Hadron Calorimeter}},
  journal       = {Sensors},
  volume        = {23},
  number        = {24},
  pages         = {9679},
  year          = {2023},
  doi           = {10.3390/s23249679},
  eprint        = {2311.04190},
  archivePrefix = {arXiv}
}

@article{Brinkerhoff:2026AutoDQM,
  author        = {Brinkerhoff, Andrew and others},
  title         = {{Anomaly Detection for Automated Data Quality Monitoring in the CMS Detector}},
  journal       = {EPJ Res. Infrastruct.},
  volume        = {10},
  pages         = {4},
  year          = {2026},
  doi           = {10.1007/s41781-025-00147-2},
  eprint        = {2501.13789},
  archivePrefix = {arXiv}
}

@article{Grosso:2023ltd,
    author = "Grosso, Gaia and Lai, Nicol{\`o} and Letizia, Marco and Pazzini, Jacopo and Rando, Marco and Rosasco, Lorenzo and Wulzer, Andrea and Zanetti, Marco",
    title = "{Fast kernel methods for data quality monitoring as a goodness-of-fit test}",
    eprint = "2303.05413",
    archivePrefix = "arXiv",
    primaryClass = "hep-ex",
    doi = "10.1088/2632-2153/acebb7",
    journal = "Mach. Learn. Sci. Tech.",
    volume = "4",
    number = "3",
    pages = "035029",
    year = "2023"
}

@article{Gavrikov:2025DINAMO,
  author        = {Gavrikov, Arsenii and García Pardiñas, Julián and Garfagnini, Alberto},
  title         = {{DINAMO: Dynamic and INterpretable Anomaly MOnitoring for Large-Scale Particle Physics Experiments}},
  journal       = {Mach. Learn. Sci. Tech.},
  volume        = {6},
  number        = {3},
  pages         = {035050},
  year          = {2025},
  doi           = {10.1088/2632-2153/ae0240},
  eprint        = {2501.19237},
  archivePrefix = {arXiv}
}

@article{Bassa:2026MEDIC,
  author        = {Bassa, Juvenal and Chattopadhyay, Arghya and Malik, Sudhir and Escabi Rivera, Mario},
  title         = {{MEDIC: a network for monitoring data quality in collider experiments}},
  journal       = {JINST},
  volume        = {21},
  pages         = {P02053},
  year          = {2026},
  doi           = {10.1088/1748-0221/21/02/P02053},
  eprint        = {2511.18172},
  archivePrefix = {arXiv}
}

@article{Belis:2024review,
  author        = {Belis, Vasilis and Odagiu, Patrick and Aarrestad, Thea Kl{\ae}boe},
  title         = {{Machine learning for anomaly detection in particle physics}},
  journal       = {Rev. Phys.},
  volume        = {12},
  pages         = {100091},
  year          = {2024},
  doi           = {10.1016/j.revip.2024.100091},
  eprint        = {2312.14190},
  archivePrefix = {arXiv}
}

@article{Kasieczka:2021LHCO,
  author        = {Kasieczka, Gregor and others},
  title         = {{The LHC Olympics 2020: a community challenge for anomaly detection in high energy physics}},
  journal       = {Rep. Prog. Phys.},
  volume        = {84},
  number        = {12},
  pages         = {124201},
  year          = {2021},
  doi           = {10.1088/1361-6633/ac36b9},
  eprint        = {2101.08320},
  archivePrefix = {arXiv}
}

@article{Aarrestad:2022DarkMachines,
  author        = {Aarrestad, T. and others},
  title         = {{The Dark Machines Anomaly Score Challenge: Benchmark Data and Model Independent Event Classification for the Large Hadron Collider}},
  journal       = {SciPost Phys.},
  volume        = {12},
  number        = {1},
  pages         = {043},
  year          = {2022},
  doi           = {10.21468/SciPostPhys.12.1.043},
  eprint        = {2105.14027},
  archivePrefix = {arXiv}
}

@article{Govorkova:2022dataset,
  author        = {Govorkova, Ekaterina and Puljak, Ema and Aarrestad, Thea and Pierini, Maurizio and Wo{\'z}niak, Kinga Anna and Ngadiuba, Jennifer},
  title         = {{LHC physics dataset for unsupervised New Physics detection at 40 MHz}},
  journal       = {Sci. Data},
  volume        = {9},
  pages         = {118},
  year          = {2022},
  doi           = {10.1038/s41597-022-01187-8},
  eprint        = {2107.02157},
  archivePrefix = {arXiv}
}

@book{Vovk:2005ALRW,
  author        = {Vovk, Vladimir and Gammerman, Alexander and Shafer, Glenn},
  title         = {{Algorithmic Learning in a Random World}},
  publisher     = {Springer},
  address       = {Boston, MA},
  year          = {2005}
}

@article{Angelopoulos:2021conformal,
  author        = {Angelopoulos, Anastasios N. and Bates, Stephen},
  title         = {{Conformal Prediction: A Gentle Introduction}},
  journal       = {Found. Trends Mach. Learn.},
  volume        = {16},
  pages         = {494--591},
  year          = {2023},
  doi           = {10.1561/2200000101},
  eprint        = {2107.07511},
  archivePrefix = {arXiv}
}

@article{Friedman:2001gbm,
  author        = {Friedman, Jerome H.},
  title         = {{Greedy function approximation: A gradient boosting machine}},
  journal       = {Ann. Statist.},
  volume        = {29},
  number        = {5},
  pages         = {1189--1232},
  year          = {2001},
  doi           = {10.1214/aos/1013203451}
}

@inproceedings{Ke:2017LightGBM,
  author        = {Ke, Guolin and Meng, Qi and Finley, Thomas and Wang, Taifeng and Chen, Wei and Ma, Weidong and Ye, Qiwei and Liu, Tie-Yan},
  title         = {{LightGBM: A Highly Efficient Gradient Boosting Decision Tree}},
  booktitle     = {Advances in Neural Information Processing Systems (NeurIPS 2017)},
  pages         = {3146--3154},
   url = {https://proceedings.neurips.cc/paper_files/paper/2017/file/6449f44a102fde848669bdd9eb6b76fa-Paper.pdf},
  year          = {2017}
}

@article{Pedregosa:2011sklearn,
  author        = {Pedregosa, F. and others},
  title         = {{Scikit-learn: Machine Learning in Python}},
  journal       = {J. Mach. Learn. Res.},
  volume        = {12},
  pages         = {2825--2830},
  url     = {http://jmlr.org/papers/v12/pedregosa11a.html},
  year          = {2011}
}

@inproceedings{Liu:2008iforest,
  author        = {Liu, Fei Tony and Ting, Kai Ming and Zhou, Zhi-Hua},
  title         = {{Isolation Forest}},
  booktitle     = {Proceedings of the 8th IEEE International Conference on Data Mining (ICDM 2008)},
  pages         = {413--422},
  publisher     = {IEEE Computer Society},
  year          = {2008},
  doi           = {10.1109/ICDM.2008.17}
}

@article{Huber:1964,
  author        = {Huber, Peter J.},
  title         = {{Robust Estimation of a Location Parameter}},
  journal       = {Ann. Math. Statist.},
  volume        = {35},
  number        = {1},
  pages         = {73--101},
  year          = {1964},
  doi           = {10.1214/aoms/1177703732}
}

@misc{uproot,
  author        = {Pivarski, Jim and others},
  title         = {{Uproot}},
  howpublished  = {Zenodo},
  doi           = {10.5281/zenodo.4340632}
}

@article{Cacciari:2008gp,
  author        = {Cacciari, Matteo and Salam, Gavin P. and Soyez, Gregory},
  title         = {{The anti-$k_t$ jet clustering algorithm}},
  journal       = {JHEP},
  volume        = {04},
  pages         = {063},
  year          = {2008},
  doi           = {10.1088/1126-6708/2008/04/063},
  eprint        = {0802.1189},
  archivePrefix = {arXiv}
}

@article{ATLAS:PFlow,
  author        = {{ATLAS Collaboration}},
  title         = {{Jet reconstruction and performance using particle flow with the ATLAS Detector}},
  journal       = {Eur. Phys. J. C},
  volume        = {77},
  pages         = {466},
  year          = {2017},
  doi           = {10.1140/epjc/s10052-017-5031-2},
  eprint        = {1703.10485},
  archivePrefix = {arXiv}
}

@article{ATLAS:TriggerRun2,
  author        = {{ATLAS Collaboration}},
  title         = {{Operation of the ATLAS trigger system in Run 2}},
  journal       = {JINST},
  volume        = {15},
  pages         = {P10004},
  year          = {2020},
  doi           = {10.1088/1748-0221/15/10/P10004},
  eprint        = {2007.12539},
  archivePrefix = {arXiv}
}

\end{document}